# Variant Plateau's Law in Atomically Thin Transition Metal Dichalcogenide Dome Networks

Boqing Liu[1†], Tanju Yildirim[2†], Tieyu Lü[3], Elena Blundo[4], Li Wang[5], Lixue Jiang[6], Hongshuai Zou[7], Lijun Zhang[7], Huijun Zhao[6], Zongyou Yin[8], Fangbao Tian[5], Antonio Polimeni[4], Yuerui Lu[1,9*]

[1]School of Engineering, College of Engineering and Computer Science, The Australian National University, Canberra, ACT 2601, Australia

[2]Center for Functional Sensor & Actuator (CFSN), Research Center for Functional Materials, National Institute for Materials Science (NIMS), 1-1 Namiki, Tsukuba, Ibaraki 305-0044, Japan

[3]Department of Physics and Institute of Theoretical Physics and Astrophysics, Xiamen University, Xiamen, 361005, China

[4]Dipartimento di Fisica Sapienza Università di Roma, 00185 Roma, Italy

[5]School of Engineering and Information Technology, University of New South Wales, Canberra, ACT 2600, Australia

[6]Centre for Catalysis and Clean Energy, Gold Coast Campus, Griffith University, Queensland 4222, Australia

[7]State Key Laboratory of Integrated Optoelectronics, School of Materials Science and Engineering, and Jilin Provincial International Cooperation Key Laboratory of High-Efficiency Clean Energy Materials, Jilin University, Changchun 130012, China

[8]Research School of Chemistry, College of Science, The Australian National University, Canberra, ACT 2601, Australia

[9]ARC Centre of Excellence in Quantum Computation and Communication Technology ANU node, Canberra, ACT 2601, Australia

[†]These authors equally contributed to this article

*To whom correspondence should be addressed: Yuerui Lu (yuerui.lu@anu.edu.au)

**Abstract:**

Since its fundamental inception from soap bubbles, Plateau's law has sparked extensive research in equilibrated states. However, most studies primarily relied on liquids, foams or cellular structures, whereas its applicability has yet to be explored in nano-scale solid films. Here, we observed a variant Plateau's law in networks of atomically thin domes made of solid two-dimensional (2D) transition metal dichalcogenides (TMDs). Discrete layer-dependent van der Waals (vdWs) interaction energies were experimentally and theoretically obtained for domes protruding in different TMD layers. Significant surface tension differences from layer-dependent vdWs interaction energies manifest in a variant of this fundamental law. Meanwhile, the remarkable mechanical properties, gas impermeability and interlayer vdWs interaction energy of TMD films enable domes and the networks to sustain high gas pressure and exist in a fundamentally variant nature for several years. Our findings pave the way towards exploring variant discretised states with applications in opto-electro-mechanical devices.

## Introduction

Since Joseph Plateau's fundamental observations[1] of soap bubbles in the 19$^{th}$ century revealed their borders meet at equal joint angles of 120°, this law has been a prominent physical law in natural science. Apart from soap bubbles[1-6], Plateau's law naturally occurs in liquid foams such as beer froth, and in solid foams such as cells, honeycombs and metallic foams[7-10]. Even in emerging two-dimensional (2D) materials such as graphene[11], the atomic arrangement of carbon atoms form a hexagonal structure at 120° to assume an equilibrated state[12]. Whilst some studies have shown that the joint angles in bubbles and foams are not always 120°, the angle deviation is quite small of only a few degrees[13-16]. Due to the low surface tension ($\sigma$) of liquids, such as 0.07-0.09 N/m for water[17] and 0.025 N/m for soap bubbles[18], the capped pressure in liquid bubbles is low and they only exist for short periods of time before bursting.

On the other hand, atomically thin layered materials such as graphene[11] and transitional metal dichalcogenides (TMDs)[19,20] are becoming promising candidate materials for the creation of nano-bubbles[21-26]. These 2D mono- and few-layer[27,28] materials are at least two orders of magnitude thinner than soap bubble films[29], but have shown strong mechanical strength[30,31], high resistance to common alkaline and acids[32,33], as well as fabulous chemical and thermal stability[34]. All previously demonstrated nano-bubbles in 2D materials were isolated singular ones without networking topologies[21-25].

Herein, we successfully created atomically thin TMD dome networks and observed variant Plateau's law in nano-scale solid systems. Variant Plateau's law is due to thickness dependent interlayer adhesion energy between the first few interacting TMD layers and the basal plane, causing thickness dependent stiffness values depending on dome layer number, resulting in large effective surface tension differences, which leads to the formation of large joint angle differences of approximately 77º between the largest and smallest angles. Results contrast with the commonly observed equal joint angles and 0º angle difference commonly associated with Plateau's law, demonstrating that this variant is observed due to our 2D TMD dome networks.

## Results and Discussion

**Generation and characterisations of TMD pressurised domes**

Recently, we successfully used a low-energy (<10 eV) proton irradiation technique to produce pressurised and spherical monolayer hydrogen domes on the surface of bulk TMDs[25]. Here, we increased the proton beam energy to ~25 eV, which allowed protons to deliberately penetrate deeper basal planes of the TMD, leading to the creation of mono-(1L), bi-(2L), and tri-layer (3L) pressurised hydrogen nano-domes (Fig. 1). Dome layer number was initially identified from optical contrast (Fig. 1a) and further confirmed by second harmonic generation (SHG) imaging performed on the same flake (Fig. 1b). The optical contrast showed a linear relationship with layer number (Fig. 1c), consistent with a previous report[35]. 1L and 3L domes exhibit intense and less intense SHG signals, respectively, and 2L domes show no SHG (Fig. 1d), which is attributed to the fact that for 2H phase TMDs, broken inversion symmetry only exits in samples with odd layer number[36,37]. The percentage yield of 2L and 3L domes was enhanced under higher proton dosage, obtained via statistical analysis (Supplementary Figure 1).

Due to the resolution limit of conventional optical microscopes, distinguishing the layer number of small-sized domes is challenging. Therefore, AFM height imaging (Figs. 2a-c) and stiffness mapping images of the domes were obtained and used to depict the layer dependent mechanical properties (Figs. 2d-f). Given the sample domes' similar radii (500-630 nm), the obtained stiffness images reveal a substantial relationship between layer number and mechanical properties. An inferred method based on a strong boundary condition (detailed in Supplementary Note 1)[38,39] was used to calculate the 2D modulus ($E_{2D}$) of the domes formed in different layers. Finite element analysis (FEA) (Supplementary Note 2) was compared against the experimental data for the nano-indentation process using the derived $E_{2D}$, illustrating good agreement (Fig. 2g). The measured stiffness showed a maximum value at the dome centre and decreased gradually by 20-30% as the AFM tip moved away from centre to the edge, which also matched with simulation results from FEA and analytically (Supplementary Figure 2). During indentation, there should be a resultant backward pressure applied due to the internal pressure encapsulated by a dome, and this may explain slight discrepancies[38,39]; however, overall, this does not appear to be a major influence as the FEA accounts for both pressures and a good agreement is found at the centre of domes. Slight discrepancies arise when moving away from the dome centre, which may be due to the non-uniform pressure distribution acting on the nano-indenter due to dome curvature. Moreover, the measured stiffness (at the dome centre) of different 1L, 2L and 3L domes exhibit an exponentially decaying trend with increasing dome radius for a constant $E_{2D}$, which agrees well with the FEA simulation as shown in Fig. 2h. FEA was also conducted for slightly modified dome footprint shapes which showed comparable stiffness values with a perfect circular dome footprint (Supplementary Figure 3 and Supplementary Note 3). The gas pressure of domes also possess an exponential trend, leading to higher pressure up to tens of MPa capped in smaller and stiffer domes (see Supplementary Figure 4). Moreover, we could estimate the pressure of domes at low $T$ to verify the inferred method against the phase transition of $H_2$ molecules, showing good agreement as in Supplementary Figure 5 and Supplementary Note 4.

Thus, using layer-dependent stiffness curves and dome radius, we can quickly identify thickness and mechanical properties of domes that match well with FEA (Figs. 2g and h). Fig. 2i reveals that the 2D modulus monolithically increases with layer number for 1-3L $WS_2$ domes, consistent with the trend reported for suspended membranes[30]. Similarly, we also used the inferred indentation method to experimentally obtain the 2D modulus of $MoS_2$ domes shown in Supplementary Figure 6a.

**Resolving layer-dependent adhesion energies in TMDs by domes**

The vdW adhesion energy ($\gamma$) plays an important role in layered materials and their physical properties, and several techniques have been reported to measure the adhesion energy between monolayer materials and the substrate[40-44], $\gamma$ in this work refers to the interlayer adhesion between TMD basal planes from the remaining bulk flake and not the base substrate. The adhesion energies in layered materials should be layer dependent, particularly for mono- and few-layer domain structures, but this layer-dependence was not previously explored. Here, our 1L, 2L, and 3L domes provide a fascinating platform to probe the intrinsic and discrete layer-dependent adhesion energies in TMDs. We firstly used density functional theory (DFT) (Supplementary Note 5) to calculate the $\gamma$ values for 1L $WS_2$ at 0 and 300 K (Fig. 3a). The calculated $\gamma$ value at 0 K is consistent with previously reported values[45]; the calculated $\gamma$ value at 300 K is ~50% higher than that at 0 K, consistent with recent results from experiments that $\gamma$ is enhanced at higher temperature[40]. The $\gamma$ values for 1L, 2L and 3L $WS_2$ were experimentally determined based on the material properties of the nano-domes (Supplementary Note 1), which showed a clear layer-dependence, matching well with our DFT calculated values at 300 K, including the presence of trapped hydrogen molecules (Fig. 3b). Minor differences may be attributed to two reasons; the first being the variation in the bond length of the topmost layers and inter-layer spacing induced by the external pressure, and thermal effects in ambient conditions[46]. Another reason may be modulated mechanical properties caused by trapped hydrogen molecules[47]. DFT results revealed that adsorbed hydrogen molecules affect the 2D modulus of monolayer TMDs, but have a small effect in few layer TMDs (Supplementary Figure 7). This explains the larger discrepancy between experimental values of adhesion energy and DFT calculated values in monolayer domes, whereas few layer domes have negligible difference. Similarly, the experimental $\gamma$ values for 1L, 2L and 3L $MoS_2$ are discrete and layer-dependent (Supplementary Figure 6b). Our $\gamma$ values for TMD monolayers (Figs. 3b and Supplementary Figure 6b) are consistent with previously reported experimental results[40].

**Identification of joint bi-dome structure and configurations**

When the exfoliated TMD thick flake is exposed to ~25 eV proton irradiation, hydrogen protons penetrate the top few basal planes of the bulk TMD flake, and then reduce to form hydrogen molecules accumulated in the top few TMD layers, forming mono- and few-layer domes (Fig. 4 a Stage I-II). With longer proton exposure time, the domes grow and interface with each other, and the inner pressure from the 1L dome (on the right) overcomes the vdW adhesion force and lifts off the top-most layer partially covering the adjacent 2L dome (on the left), eventually forming a joint double dome (bi-dome) system (Fig. 4 a Stage III-V). Therefore, the hydrogen molecules in our 1L, 2L and 3L domes exist in different basal planes of the TMD, leading to intriguing different joint dome configurations and interaction dynamics among domes. It is important to note here that the substrate is the TMD flake beneath the resulting domes and not the base $SiO_2$/Si substrate, as growing domes on different base substrates still results in a universal height to radius ratio[25]. Experimentally, many different coalescing dome formations were observed, and the most common case was the bi-dome presented in the AFM image as shown in Fig. 4b. This case is the existence of a bi-dome composed of a large 1L dome coinciding with a smaller 2L dome, demonstrating complex interaction between hydrogen accumulation underneath different basal planes of the TMD. To maintain stability, domes sharing common vertices must be in different layers of the TMD; otherwise, domes will merge resulting in larger domes in the same basal plane to minimise their surface energy, analogous to liquid bubbles. The structure and configuration of a joint bi-dome was directly confirmed by a bursting experiment test (Supplementary Figure 8 and Supplementary Note 6). To further characterise the structure and confirm the configuration, stiffness mapping was conducted among the joint domes and a clear dependency between the stiffness of interacting domes was observed (Fig. 4c). This reaffirms domes are in different layers as illustrated by the inset in Fig. 4c, where the smaller 2L dome exhibits much higher stiffness than the larger 1L dome. FEA simulations of the generated height profile and stiffness of the two merging domes (with configuration depicted by the inset in Fig. 4c) is given in Figs. 4d and e, respectively, illustrating good agreement between the simulation and experimental results. Some discrepancies near the transition region are visible, which is attributable to the AFM resolution or a mismatch in modelling the geometry and boundary conditions. This demonstrates that domes function as highly pressurised membranes with high tension across the dome interface between internal and external pressures (Supplementary Note 7). Moreover, a very sharp stiffness transition edge between dome layers is visible (Fig. 4e), further confirming the two domes are sitting in different basal planes. The same method was applied to a different joint configuration with a larger 2L region and smaller 1L region, also showing good agreement between simulation and experiment (Supplementary Figure 9).

**Observation of variant Plateau's law in joint dome systems**

In a bi-dome system, three joint angles between adjoining edges were formed, termed as $\alpha_1$, $\alpha_2$ and $\alpha_3$ (inset in Fig. 5a). The values of $\alpha_1$, $\alpha_2$ and $\alpha_3$ for $WS_2$ bi-domes were measured to be 153.1º ± 8.8º, 75.6º ± 9.8º and 131.2º ± 6.1º, respectively (Fig. 5a), differing with each other, which is in great contrast to Plateau's law of equal joint angles of 120° as observed in soap bubbles[1]. The values of $\alpha_1$, $\alpha_2$ and $\alpha_3$ for $WS_2$ bi-domes were calculated (Supplementary Note 8 and Supplementary Figure 10) to be 152.6º, 72.3º and 135.1º, respectively, agreeing well with the experimental values (Fig. 5a). A large difference of ~77.5º between $\alpha_1$ and $\alpha_2$ was observed in $WS_2$ bi-dome systems, such large deviation has not been previously reported, demonstrating a variant Plateau's law. The governing reason for this behaviour is due to the large unequal effective surface tension ($\sigma$) differences in the different dome layers, where 1L, 2L and 3L domes have experimental $\sigma$ values of 2.4 ± 0.3, 3.5 ± 0.3, 3.6 ± 0.3 N/m, respectively, which is significantly higher than other systems (Supplementary Note 8). The large difference among the $\sigma$ values of 1L, 2L and 3L domes is a direct result of the discrete nature of $\gamma$ in 1-3L TMDs as shown in Fig. 3b. Additionally, some domes in this work are partially sitting on top of others; the dome part sitting on top of another dome has different $\gamma$ compared to the part being exfoliated from the bulk flake. For example, it requires less energy to detach a 1L sheet from a 2L sheet, rather than from the bulk (Supplementary Figure 10b). Domes merging in this manner demonstrate the discrete nature of $\gamma$, leading to significant variations in the effective $\sigma$ of each dome. Similarly, large variations of the joint angles were also observed and analytically calculated for MoS$_2$, demonstrating excellent agreement (Supplementary Figure 11a).

To exacerbate the large deviations from Plateau's law in TMD domes, the joint angles $\beta_1, \beta_2$ and $\beta_3$ of WS$_2$ tri-dome systems were also experimentally determined as shown in Figs. 5b-d. The values of $\beta_1, \beta_2$ and $\beta_3$ for WS$_2$ tri-domes were measured to be 135.5º ± 6.4º, 104.6º ± 7.1º and 119.9º ± 3.7º, respectively, also clearly demonstrating variant Plateau's law, whereby joint angles no longer merge at 120º. Similar large angle deviations were also observed in MoS$_2$ tri-domes (Supplementary Figure 11b). The largest angle was found in the 1L region in a tri-dome system, similar to that observed in a bi-dome system; the angle difference ($\beta_1$ - $\beta_2$) in a tri-dome system is smaller than that in a bi-dome system ($\alpha_1$ - $\alpha_2$). A statistical analysis of angle variation in WS$_2$ and MoS$_2$ bi-dome and tri-dome networks (Figs.5a and b, and Supplementary Figure 11) concluded that $\alpha_1$ and $\beta_1$ for both material systems (WS$_2$ and MoS$_2$) are the largest among all three angles. Moreover, bi-dome systems have larger angle deviation compared to tri-dome systems. The $\alpha_1$, $\alpha_2$ and $\alpha_3$ of a WS$_2$ bi-dome system deviate by 8.8º, 9.8º and 6.1º, respectively, whilst $\beta_1, \beta_2$ and $\beta_3$ exhibit lower deviations of 6.4º, 7.1º and 3.7º, respectively. This could be attributed to the fact that bi-domes are more likely to be influenced by the surrounding conditions on the bulk flake, including side domes, surface contaminants and surface topology, leading to larger deviation in joint angles $\alpha$. Owing to a trigonal structure in tri-dome systems, a better structural stability generates smaller deviation in joint angles compared to bi-domes, and WS$_2$ and MoS$_2$ tri-dome systems have comparable angle values for $\beta_1, \beta_2$ and $\beta_3$.

In addition, it is worth to note that the validity of the variant of Plateau's law in complex dome network systems. To prove this, we characterised a one-dimensional (1D) dome chain and 2D superlattice structure with bi-dome and tri-domes as the repeating unit (Supplementary Figure 12). In the 1D dome chain, it also proves that the largest and smallest angles are $\alpha_1$ in 1L regions and $α_2$ in 2L regions, respectively, which is confirmed by stiffness mapping (Supplementary Figure 12b). This reaffirms that the variations in joint angles, regardless of the complexity of the network structure, are induced by layer-dependent mechanical properties including $E_{2D}$, $γ$ and $σ$. Furthermore, due to the robustness of TMDs, $E_{2D}$ and $γ$ remain constant over time, allowing the morphological structure and joint angles to endure for a long period without variations (Supplementary Figure 13), even in the presence of external stimuli (nano-indentation and optical tests). This means determination of joint angles of the network is a reliable tool to quickly identify layer number of domes in a network, exemplified by the 2D dome network shown in Supplementary Figure 12c.

We successfully fabricated 1L, 2L and 3L pressurised hydrogen nano-domes, and their networks, using low-energy proton irradiation on TMD flakes. Dome layer number was identified and confirmed by optical contrast, SHG imaging and high-resolution stiffness mapping. The layer-dependent mechanical properties of domes revealed by AFM nano-indentation match well with FEA simulations. 1-3L domes provide a fascinating platform for resolving the discrete layer-dependent adhesion energies in TMDs, which matched with DFT calculations. Joint bi-dome and tri-dome systems were formed when the domes sitting in different basal planes grew and interacted with each other. The layer-structure and configurations of joint domes were successfully identified and confirmed by stiffness mapping, FEA calculation and bursting test. Variant Plateau's law was observed in bi-dome and tri-dome systems, which is due to the large variations of effective surface tension in different layer number domes, arising from the discrete vdWs adhesion energy in TMDs. For example, our bi-dome system has a large joint angle difference of more than 77º, in great contrast to conventional liquid bubble systems that have equal joint angles. Our solid domes equivalent surface tension values range from 2.4-3.6 N/m, which is two orders of magnitude larger than their liquid or foam counterparts enabling high pressure encapsulation. The variant Plateau's law observed in experiments agrees with our analytical calculation. Furthermore, stable and long-lasting 1D and 2D dome networks were realised in experiments with multiple domes sitting in different basal planes of the TMD. Dome networks can become a promising topic for generating 1D or 2D nanostructures with large variation in properties enabling new device applications in the fields of nano-photonics, nano-opto-mechanics and quantum science.

## Methods

**Fabrication of the domes**

Thick TMD flakes are mechanically exfoliated onto various substrates including $SiO_2$/Si (275 nm $SiO_2$), Au deposited and doped silicon. The samples are subsequently placed in the chamber for proton irradiation treatment in a high vacuum condition. During the treatment, the pressure inside the chamber is P = $1.4 \times 10^{-4}$ mbar, with hydrogen gas delivered at a flow rate of 30 sccm. Meanwhile, the $H^+$ protons produced in an ionisation chamber will be accelerated and guided onto the sample surface, as a form of proton beam with energy at least 25 eV. The relative larger energy enables protons to deliberately penetrate deeper basal planes of the TMD, creating different layered domes and dome networks. The entire treatment process normally takes a few hours to complete.

**Optical characterisation**

Optical microscope images were taken by a Zeiss 780 confocal microscope equipped with a 633 nm single photon laser. SHG measurement were also performed on a Zeiss 780 confocal microscope using a Ti:Sapphire laser (150 fs, 80 MHz). The sample is excited and measured under a 50× confocal objective lens (NA= 0.85), and results are collected in the reflection mode at a fundamental laser wavelength of 900 nm. All data of optical characterisation measurements were processed and analysed by image processing software, Zen 3.2 (blue edition).

**AFM Nano-indentation**

The topographic images were captured using a Bruker Dimension Icon AFM. The aspect ratio of the domes was measured in Scanasyst mode with soft cantilevers, Scanasyst-Air, whose nominal spring constant $k$ and nominal tip radius are 0.4 N/m and 2 nm, respectively. Using soft tips to obtain the profile and topographic information is designed to minimise tip–sample interaction and avoid modifying the shape of the domes or destroy domes. Indentation experiments on TMD domes were performed in Quantum Nano Mechanical (QNM) mode. The tips used for the measurement is RTESPA-300 with $k$ = 40 N/m and a nominal $R_{tip}$ = 8 nm. Before the measurement, the tips required calibration in terms of resonance frequency, deflection sensitivity, contact area, and spring constant to ensure results are repeatable. The indentation force was adjusted and set between 50 to 100 nN to avoid the fact that large forces might destroy domes. For the force curves at different locations, they were extracted using ramp mode and this mode could record extending and extracting cycles at different locations on the dome samples. Stiffness values could be processed by calculating the slope of the loading part of extending cycles. For the adhesion energy measurement, the 2D modulus ($E_{2D}$) was extracted and calculated based on the force curves at the centre of the dome samples.

**Data availability**

The authors declare that all data supporting the finding could be found in the manuscript and supporting information of this work. The datasets generated during the current study are available upon request from the corresponding author.

## Supplementary Information

All additional data and supporting information and methods are presented in the supplementary information file available online. The figures and information in the supplementary information has been cited at appropriate places in this manuscript.

## Acknowledgements

The authors acknowledge funding support from ANU PhD student scholarship, Australian Research Council (ARC; numbers DP220102219, DP180103238, LE200100032) and ARC Centre of Excellence in Quantum Computation and Communication Technology (CE170100012), and National Heart Foundation (ARIES ID: 35852). T.Y. acknowledges the Applied Mechatronics and Biomedical Engineering Research (AMBER) group at the University of Wollongong, Australia, for their COMSOL computing resources where he stayed for a small part during the preparation of this work. B. L. would like to acknowledge the support from Centre of Advanced Microscopy (CAM), Australian National University. The authors would like to acknowledge the helpful discussion with Professor Rui Huang from the University of Texas at Austin.

## Author Contributions

Y. L. conceived and supervised this study; B. L., T. Y. and Y. L. designed the experiments; B. L. exfoliated TMD flakes; E. B. and A. P. conducted proton irradiation; B. L. conducted AFM nano-indentation experiments, performed the optical characterization and SHG measurements; B. L. and T. Y. contributed to schematics; T. L. conducted DFT calculations; T. Y. conducted FEA calculation, fitting and simulation; L. J., H. J. Z. and Z.Y. provided facility support; H. Z., L. W., L. Z and F. T. provided simulation support; B. L., T. Y. and Y. L. analysed the data and drafted the manuscript; and all authors participated in manuscript editing and discussions.

## Competing interests

The authors declare that they have no competing interests.

**Captions**

**Figure 1 | Optical characterisation of mono- and few-layer pressurised nano-domes in TMDs. a**, Optical microscope image of pressurised domes generated on a $WS_2$ flake, whereby mono- (1L), bi- (2L) and tri-layer (3L) domes have been highlighted by blue, red and green circles, respectively. The layer number of the dome is initially identified from the optical contrast. **b**, Second harmonic generation (SHG) mapping of the same flake shown in **a**. 1L domes show the strongest SHG signal, whilst 2L domes are almost invisible under the same excitation and collection conditions. **c**, Measured optical contrast as a function of dome layer number. The measured optical contrast for 1-3L domes with similar sizes are 11.9% ± 1.3% (blue), 22.0% ± 0.9% (red) and 31.9% ± 2.0% (green), respectively. The error bars represent statistical variation from at least 4 domes for each group with different layer number. **d**, Measured SHG intensity as a function of layer number of the domes. The measured SHG intensities for 1-3L domes with similar sizes are 36.93 ± 3.74, 0.54 ± 0.01 and 9.79 ± 3.36 arbitrary unit (a.u.), represented by blue, red and green bars respectively. The error bars represent statistic variation from at least 5 domes from each group with different layer number.

**Figure 2 | Mechanical characterisations of mono- and few-layer pressurised TMD nano-domes. a-c**, Atomic force microscope (AFM) images of a 1L (**a**), 2L (**b**) and 3L dome (**c**). The measured centre dome height ($h_m$) to radius ($R$) ratio ($h_m/R$) for 1-3L domes are 0.18, 0.17 and 0.16, respectively. **d-f**, Stiffness mapping images measured for the 1-3L domes shown in **a-c**. **g**, Force-indentation curves (dots) measured at the centre of domes with different layer numbers (1-3L). The simulated force-indentation curves (solid lines) from finite element analysis (FEA) match very well with the measured ones. The inset shows the experimental set up for data acquisition: an AFM tip was used for the mechanical nano-indentation of domes. **h**, Measured stiffness as a function of dome radius (solid dots), for 1-3L domes. Simulated results (lines) by FEA match well with experimental observations. **i**, Extracted two-dimensional modulus ($E_{2D}$) of $WS_2$ as a function of layer number. The extracted $E_{2D}$ values are 244.0 ± 35.7, 429.5 ± 37.7 and 520.4 ± 49.2 N/m for 1-3L $WS_2$ domes, respectively. The error bars represent statistical variation from 20 domes for each group with different layer number.

**Figure 3 | Resolving layer-dependent adhesion energies in TMD by nano-domes. a**, Calculated energy required to separate a $WS_2$ monolayer from the bulk as a function of separation distance ($d$), by DFT (dots) and analytical solution (line), at 0 (blue) and 300 K (red). $d_0$ is the initial separation distance. The adhesion energy is defined to be the maximum energy

required to separate the layer from the bulk. The adhesion energies of a $WS_2$ monolayer at 0 and 300 K are calculated to be 21.5 and 29.7 meV/Å$^2$, respectively. The analytical solution has the form $\gamma_{dist} = \gamma \left[\frac{3}{2}(\frac{d_0}{d})^3 - \frac{1}{2}(\frac{d_0}{d})^9\right]$, where $\gamma_{dist}$ is the monolayer-surface interaction energy per unit area at a corresponding separation distance and $\gamma$ is the interfacial adhesion energy per unit area[45]; The inset provides the schematic illustration of the initial interlayer distance ($d_0$) and modified separation distance ($d$) in $WS_2$. **b**, Measured layer-dependent adhesion energies of $WS_2$, which are the energies required to exfoliate top layers (1-3L) from a $WS_2$ bulk flake, using nano-domes. The adhesion energy values obtained from 1-3L $WS_2$ are 33.8 ± 3.0, 42.6 ± 2.5 and 45.0 ± 4.2 meV/Å$^2$, all of which are in good agreement with the DFT results of 31.5, 40.3 and 43.5 meV/Å$^2$, respectively. The error bars represent statistical variation from 20 domes for each group with different layer number.

**Figure 4 │ Identification of structure configuration of a joint bi-dome *via* nano-indention. a.** Schematic illustration of a joint bi-dome formation process. **b**, AFM image of a $WS_2$ joint bi-dome made of a large 1L dome and small 2L dome. **c,** Stiffness mapping of the joint bi-dome shown in **a**. The inset indicates the structure of the domes along the white dashed line in **b**. **d**, Measured height profile (grey) of the joint bi-dome along the white dashed line shown in **b**. The simulated profile (red) generated by FEA reasonably matches with the measured one. **e**, Measured stiffness (solid blue dots) as a function of scan distance along the white dashed line shown in **b**. Simulated stiffness values by FEA calculation (red diamond) well match with experimental values. The sharp drop of stiffness value at 300 nm corresponds to the joint boundary between 2L and 1L domes.

**Figure 5 │ Observation of variant Plateau's law in bi- and tri-dome systems. a**, Histogram of the joint angles ($\alpha_i$, $i$ = 1, 2, 3) in $WS_2$ bi-dome systems, extracted experimentally (solid bars) and analytically (patterned bars). The inset shows the angle notations in a standard bi-dome configuration. The experimental values for joint angle are 153.1° ± 8.8°, 75.6° ± 9.8° and 131.2° ± 6.1° for $\alpha_1$, $\alpha_2$ and $\alpha_3$, respectively. Statistical data was collected from 20 joint bi-domes. The calculated values for $\alpha_1$, $\alpha_2$ and $\alpha_3$, are 152.6°, 72.3° and 135.1°, respectively, all of which are in good agreement with experimental values. **b**, Histogram of the joint angles ($\beta_i$, $i$ =1, 2, 3) in $WS_2$ tri-dome systems, extracted experimentally. The inset shows the angle notations in a standard tri-dome configuration. The experimental values for joint angle are 135.5° ± 6.4°, 104.6° ± 7.1° and 119.9° ± 3.7° for $\beta_1$, $\beta_2$ and $\beta_3$, respectively. Statistical data was collected from 15 joint tri-domes. **c**, AFM image of a $WS_2$ tri-dome system, consisting of 1L, 2L and 3L

domes joint together. **d**, Stiffness mapping image of tri-dome shown in **c**. The layer-dependent stiffness can be clearly distinguished.

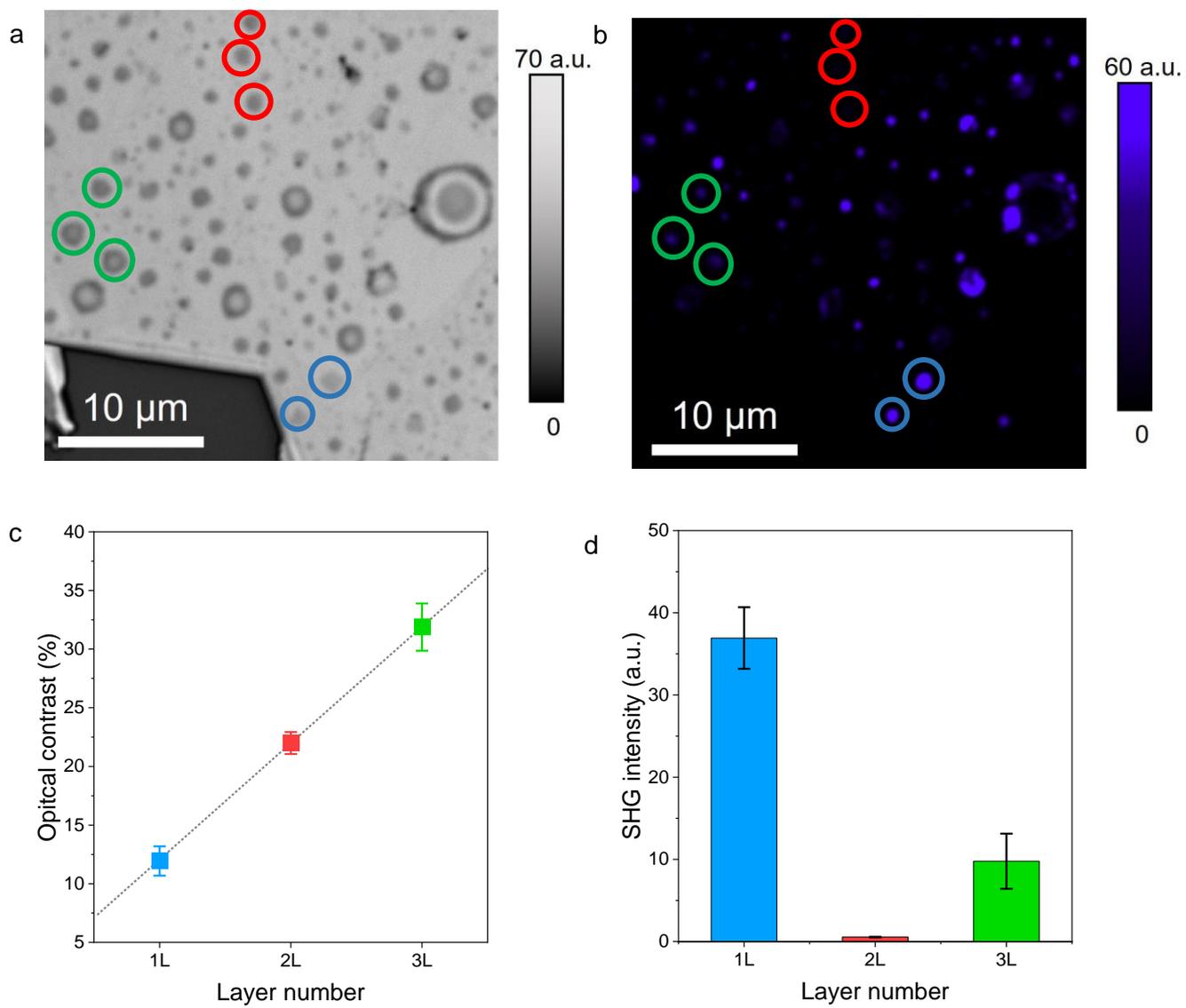

**Figure 1**

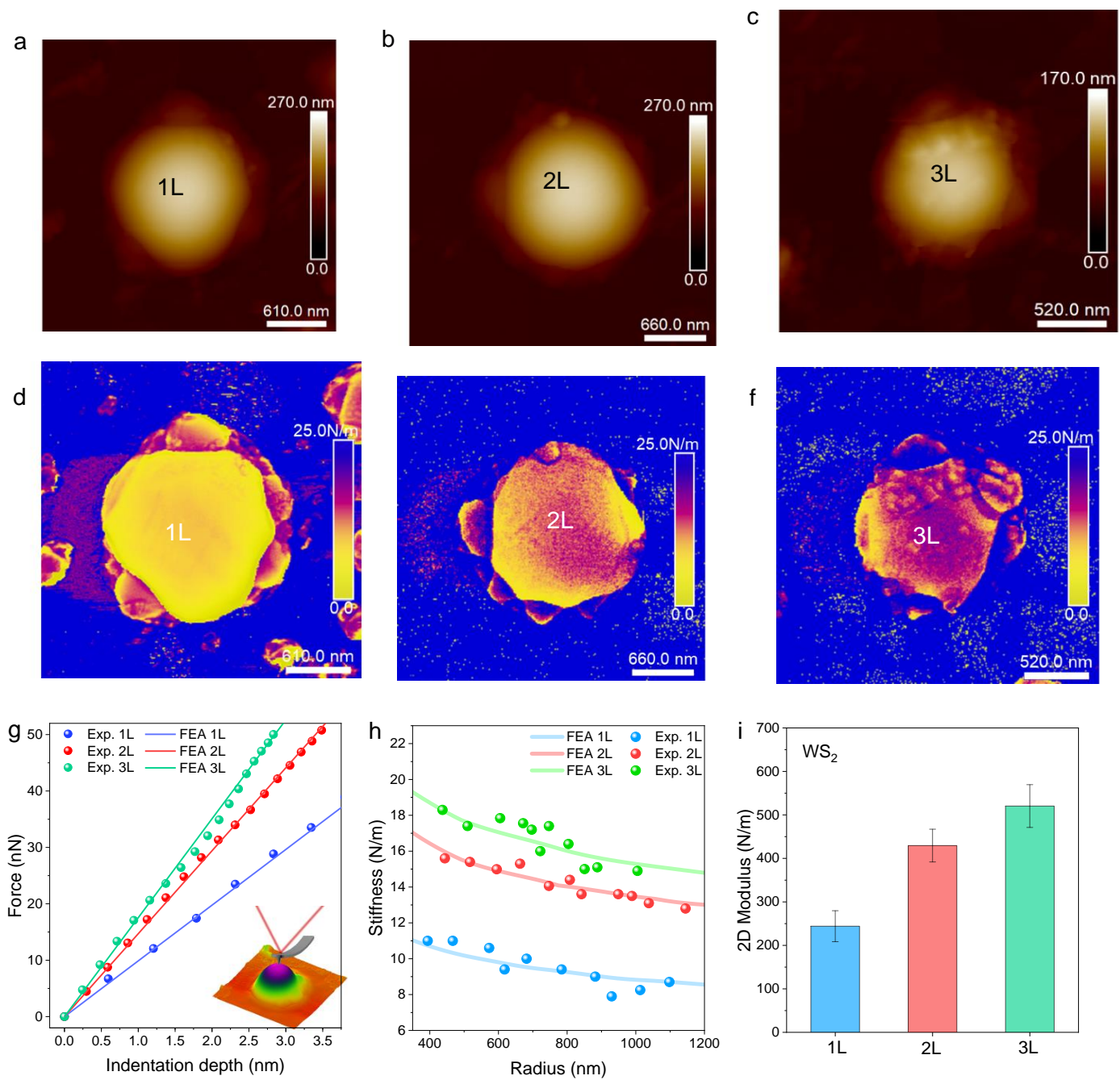

**Figure 2**

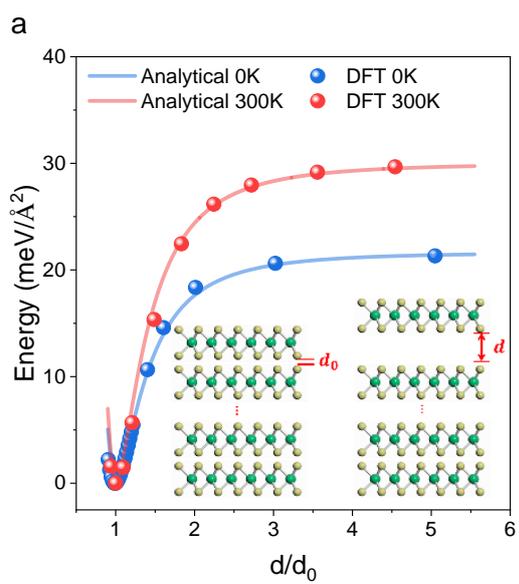 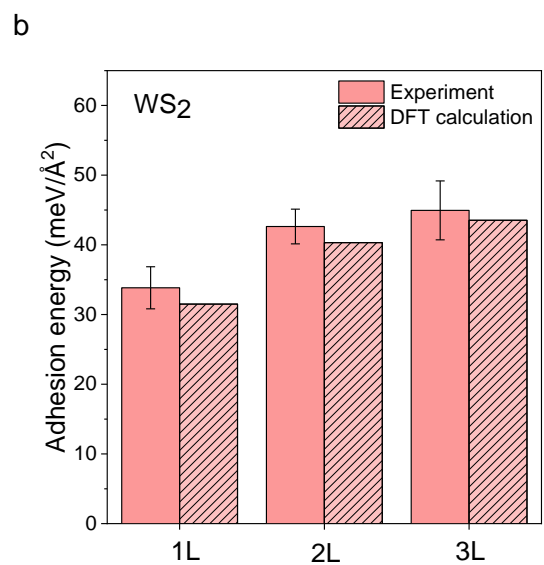

**Figure 3**

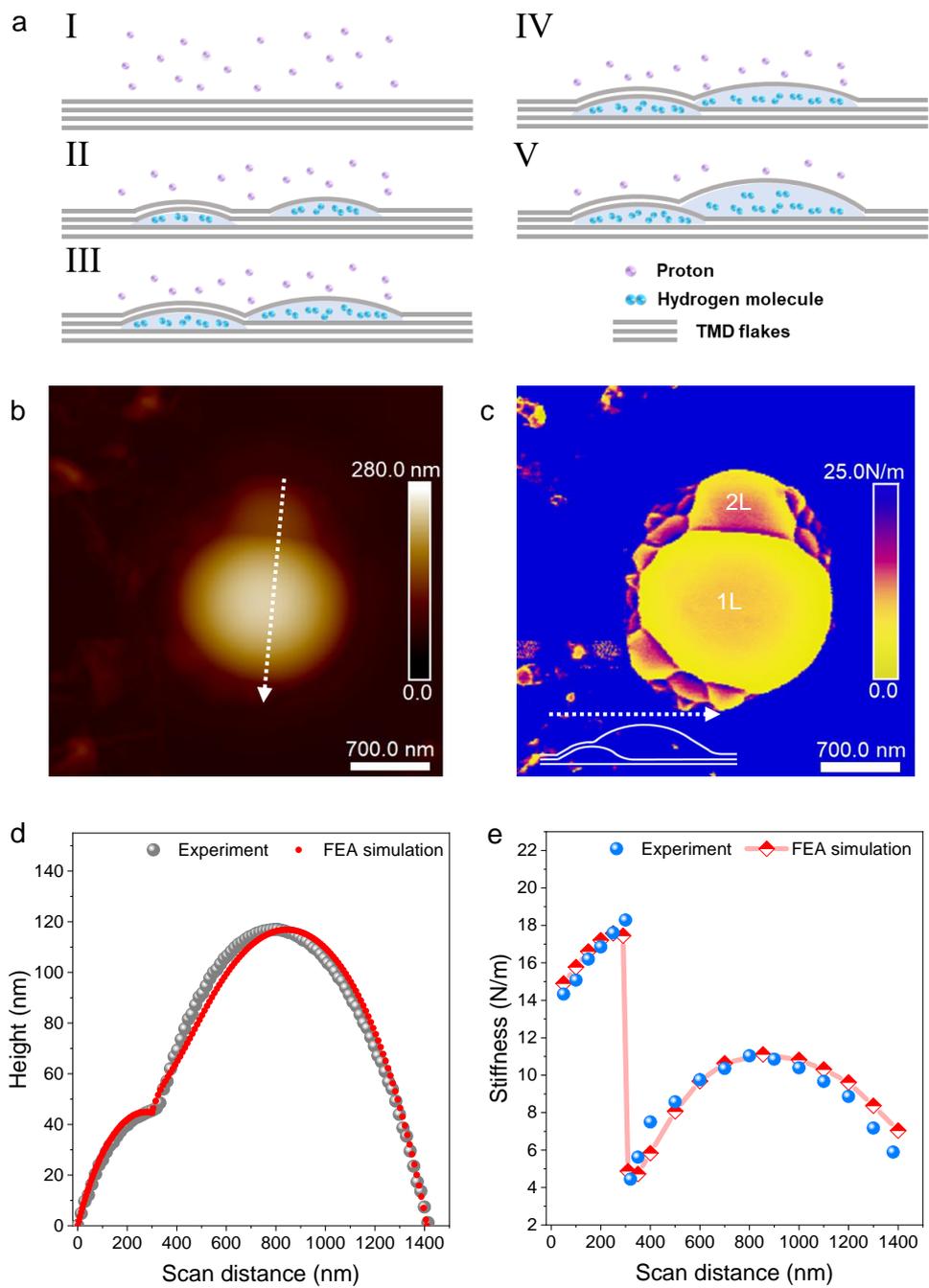

**Figure 4**

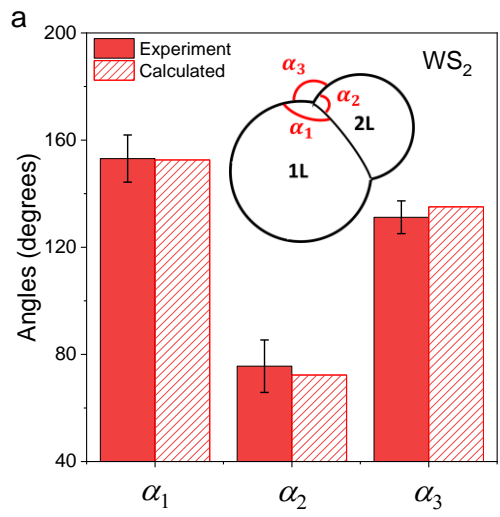
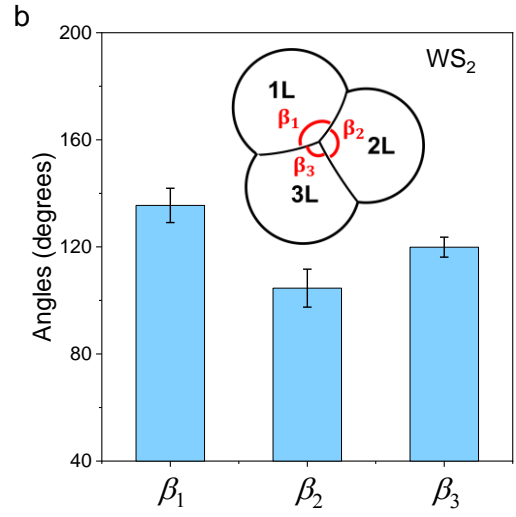
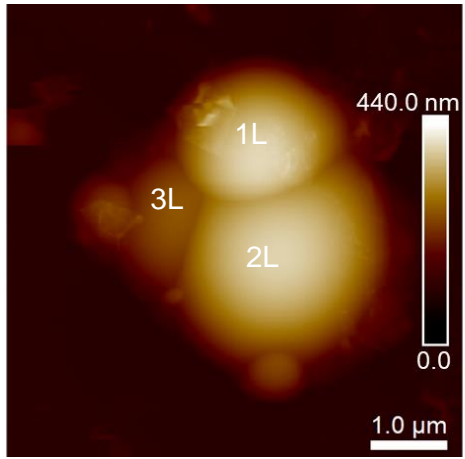
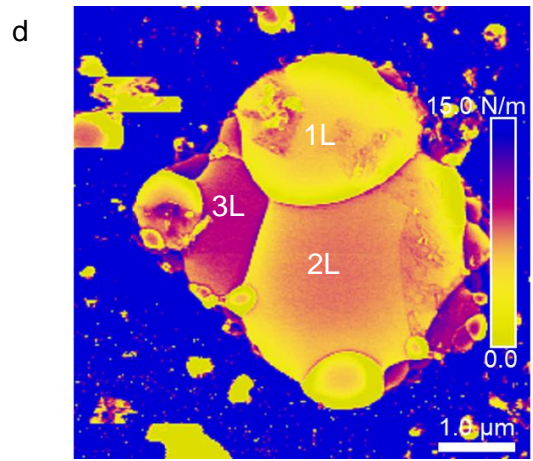

**Figure 5**